\documentclass[12pt]{iopart}

\newcommand{\mb}[1]{\mbox{\boldmath $#1$}}
\newcommand{\npb}{$\{\mb{k},\mb{\ell},\mb{m},\mb{\bar{m}}\}\;$}
\def \ts {\textstyle}

\begin{document}

\jl{6}

\title[General approach to the study of vacuum space-times with an
isometry]
{General approach to the study of vacuum space-times with an isometry}

\author{Francesc Fayos\dag$\S$\ and Carlos F. Sopuerta\ddag
\footnote[3]{Also at the Laboratori de F\'{\i}sica Matem\`atica,
Societat Catalana de F\'{\i}sica, I.E.C., Barcelona, Spain}}

\address{\dag\ Departament de F\'{\i}sica Aplicada, UPC,
E-08028 Barcelona, Spain}

\address{\ddag\ Relativity and Cosmology Group, School of Computer Science
and Mathematics, Mercantile House, Hampshire Terrace, PO1 2EG Portsmouth,
United Kingdom}

~

\address{E-mail: {\tt labfm@ffn.ub.es}\,, {\tt carlos.sopuerta@port.ac.uk}}

\begin{abstract}
In vacuum space-times the exterior derivative of a Killing vector field
is a 2-form (named here as the Papapetrou field) that satisfies
Maxwell's equations without electromagnetic sources.  In this
paper, using the algebraic structure of the Papapetrou field, we will
set up a new formalism for the study of vacuum space-times with an
isometry, which is suitable to investigate the connections between the 
isometry and the Petrov type of the space-time.  This approach has some 
advantages, among them, it leads to a new classification of these 
space-times and the integrability conditions provide expressions that 
determine completely the Weyl curvature.  These facts make the formalism 
useful for application to any problem or situation with an isometry and
requiring the knowledge of the curvature.
\end{abstract}




\section{Introduction}
The search of exact solutions of the Einstein field equations
has been one of the most active fields of research in general
relativity.  Despite of the non-linearity of the Einstein
equations and its invariance under general changes of coordinates,
a wide variety of exact solutions have been found (see~\cite{KSHM}).
Although we do not have exact solutions describing the
gravitational field in complex realistic systems,
the development of techniques for the search of exact solutions
as well as the study of some particular ones have been of crucial
importance for the understanding of general relativity and for some
developments within its framework, like perturbation theory, numerical
methods, etc.
There are lots of techniques and methods to find exact solutions
of Einstein's equations.  Among them, we want to emphasize two:
First, the imposition of symmetries (Killing symmetries, conformal Killing
symmetries, etc.) on the space-time metric and second, the
imposition of a special algebraic structure for the space-time
(described by the algebraic structure of the Weyl tensor).
It is worth to note that many important
techniques have been developed starting from these two types
of simplifications of the Einstein equations.   However,
despite of the number of works dedicated to these techniques
and their applications, as far as we know there are few studies
establishing connections
between the existence of symmetries and the particular algebraic
structure of the space-time, even in the case of Killing symmetries
(see~\cite{KSHM}, chapter~33).
In this paper we will set up a new formalism to study solutions
of Einstein's equations with a Killing vector field (KVF hereafter)
which, as we will see, is suitable to study such connections as
well as other important related subjects.  Such a formalism is
well-motivated taking into account that there is a great variety of
physical situations with a symmetry described by a KVF.

The starting point is the fact, firstly noticed by Papapetrou~\cite{PAPA},
that the exterior
derivative of a KVF is a 2-form satisfying Maxwell's equations,
with the KVF playing the role of the electromagnetic
potential and satisfying the covariant Lorentz gauge.
This 2-form, which we will call the {\em Papapetrou} field
associated with the KVF (sometimes also called the Killing 2-form or
the Killing bivector~\cite{DEB1,DEB2}), has been used and applied to several
subjects, like the search and study of exact solutions of Einstein's field
equations (see, e.g.,~\cite{HOMI,CAMI}), or the study
of black holes in the presence of external electromagnetic fields
(see~\cite{WALD}). Other applications were described in~\cite{FASO},
where a systematic study was made.  Specifically,  general covariant
expressions for the principal null direction(s) of the Papapetrou field
in terms of quantities associated with the Killing were found, and
moreover, some differential properties of the principal null
direction(s) were studied and the conditions for a principal direction
to be aligned with a multiple principal direction of an algebraically
special vacuum space-time were given.

The existence of the Papapetrou field provides a classification
of the vacuum\footnote{For the sake of simplicity we will restrict
ourselves to vacuum space-times, but the main ideas
on which our formalism is based can be extended to space-times with
other types of energy-momentum content.} space-times having a KVF.
This classification will consider whether the Papapetrou field
is regular (two different null principal directions) or singular
(only one null principal direction), the  Petrov type, and the
possible different alignments of the principal directions of the
Papapetrou field with those of the space-time (the principal directions
of the Weyl tensor).  This classification will lead us to consider
a new approach based on the well-known Newman-Penrose (NP hereafter)
formalism~\cite{NEPE} and on the structure and properties of the
Papapetrou field.  The first important point is that we will write
all the equations involved using a NP basis adapted to the principal
direction(s) of the Papapetrou field.
And the second important point is that we will extend the
usual framework of the NP formalism by adding new variables and their
corresponding equations.  The first set of new variables will be
the components of the KVF in the adapted basis chosen, and the equations
that we will use for them are the equations
defining the Papapetrou field combined with the Killing equations.
To close the system we will need to add two new variables: The eigenvalues
in the regular case, or a complex function in the singular case.  
The equations for these two
variables are just the Maxwell equations for the Papapetrou field.
All these ingredients together provide a framework that incorporates
explicitly the existence of a KVF and that allows to control in a
clear and transparent way the algebraic structure of the space-time
and the possible alignments of the principal direction(s) of the
Papapetrou field with those of the Weyl tensor.

Using this framework, we have studied the integrability conditions for
some variables and the compatibility conditions of other sets of
equations.  As a result of this study we have obtained
explicit expressions for the components of the Weyl tensor in
terms of the components of the KVF, some spin coefficients,
and the two additional functions associated with the Papapetrou field.
Therefore, it is not necessary to solve the second Bianchi
identities for them, instead these identities will provide more
equations for the spin coefficients (apart from the NP equations).
As we will see,  other integrability or compatibility conditions
will be also reduced to equations for the spin coefficients.
To sum up, this formalism provide expressions for the Weyl tensor
(which depend only on spin coefficients and variables associated
with the KVF) and the main problem is reduced to study the set
of equations for the spin coefficients (the NP equations and
the integrability and compatibility conditions).
These characteristics of the formalism make it very appropriate
for any situation or problem requiring the knowledge of the
curvature (the Weyl tensor) and its properties.

The plan of the paper is as follows: In section~\ref{alge} we review
some useful material on the algebraic structures of the Papapetrou field
and the Weyl tensor, and we will propose a new classification for 
vacuum space-times with a KVF.
In section~\ref{form} we introduce the formalism for the case of vacuum
space-times.  In section~\ref{inte} we study in full generality the
integrability and compatibility conditions for some sets of equations
and describe a general scheme to proceed further. In section~\ref{exam}
we apply our formalism to two particular examples.  First, we will study 
completely the case in which the Papapetrou field associated with the KVF 
is singular.  And in the second example, we will examine the case of 
Petrov type III vacuum space-times, specifically we will study the
possibility of alignment of the multiple principal direction of the
space-time with one of the Papapetrou field.  We will finish with some
comments and remarks in section~\ref{core}.  In~\ref{appa} we give some
formulae for the Kundt class of Petrov types N and III vacuum metrics
that have been used in the examples of section~\ref{exam}.
Through this paper we will follow the notation and conventions
of \cite{KSHM} unless otherwise stated.

\section{Some important facts and ideas\label{alge}}

In order to introduce a new formalism to deal with vacuum space-times
with a KVF, as the one described in the introduction, it will be crucial
to consider algebraic structures associated (in a local way) with the KVF
and with the space-time.
In the case of the space-time the algebraic structure mostly used and
studied in the literature is the algebraic structure of the Weyl tensor,
whose classification (nowadays called Petrov
classification~\cite{PETR,BEL1,BEL2}) can be found, for instance,
in~\cite{KSHM,DINV,STEP,CHAN}.
On the other hand, in the case of a KVF,
the algebraic structure to be associated with it will be the algebraic
structure of its exterior derivative.  Since the exterior derivative
of a KVF is a 2-form, the algebraic classification will be identical to
that of the electromagnetic field.  In what follows we will review
these algebraic structures and introduce a new classification for
space-times with a KVF, establishing some notation and introducing
some formulae that will be used along this paper.

One of the starting points in the development of this work is the fact,
already recognized by Papapetrou~\cite{PAPA}, that a KVF $\mb{\xi}$
can always be seen as the
vector potential of an electromagnetic field, the Papapetrou field
associated with $\mb{\xi}$, defined as follows
\begin{equation}
\mb{F}\equiv\mb{d\xi} \,, \label{papa}
\end{equation}
where $\mb{d}$ denotes the exterior derivative.  Using the
Killing equations
\begin{equation}
\xi_{a;b}+\xi_{b;a}=0 \,, \label{kill}
\end{equation}
where a semicolon means covariant differentiation, and the Ricci
identities for $\mb{\xi}$
\begin{equation}
\xi_{a;bc} = R_{abcd}\xi^d \,, \label{rici}
\end{equation}
where $R_{abcd}$ denotes the Riemann tensor, we can show that
$F_{ab}$ ($=\xi_{b;a}-\xi_{a;b}=2\xi_{b;a}$) satisfies the Maxwell
equations
\begin{equation}
F_{[ab;c]}=0, \quad F^{ab}{}_{;b}= J^a \,, \label{maxe}
\end{equation}
where $J^a$ is the conserved current given by
\[  J^a\equiv 2R^a{}_b\xi^b ~~ \Rightarrow ~~
J^a{}_{;a} = 0 \,, \]
where $R_{ab}\equiv R^c{}_{acb}$ is the Ricci tensor. It follows
directly from these expressions that if the KVF is an
eigenvector of the Ricci tensor with zero eigenvalue, then $F_{ab}$
satisfies Maxwell's equations in the absence of electromagnetic charge
and current distributions.  Obviously, this is in particular true for
vacuum space-times.  Finally, as a consequence of the Killing
equations~(\ref{kill}), $\mb{\xi}$ satisfies the covariant version of
the Lorentz gauge
\[ \xi^a{}_{;a}=0 \,. \]

From now on we will consider that the algebraic structure of the
Papapetrou field $F_{ab}$~(\ref{papa}), a 2-form, is the algebraic
structure associated with the KVF $\mb{\xi}$.  In an arbitrary NP basis
\npb$\!$, a basis made up of two real null vectors $(\mb{k},\mb{\ell})$
and a complex null vector $\mb{m}$ and its complex conjugate
$\mb{\bar{m}}$ (a bar means complex conjugation) such that
\[ \mb{k}\cdot\mb{\ell}=-\mb{m}\cdot\mb{\bar{m}}=-1\,, \hspace{5mm}
\mb{k}\cdot\mb{m}=\mb{\ell}\cdot\mb{\bar{m}}=0\,, \]
an arbitrary 2-form $F_{ab}$ ($F_{[ab]}=F_{ab}$) can be written in
the following form
\[ \tilde{F}_{ab} = \Phi_0 U_{ab}+\Phi_1 W_{ab}+\Phi_2 V_{ab} \,,\]
where the tilde denotes the self-dual operation
\[ \tilde{F}_{ab} = F_{ab} + i*F_{ab} \hspace{1cm}
(*F_{ab}\equiv\textstyle{1\over2}\eta_{ab}{}^{cd}F_{cd})\,, \]
$*$ denotes the dual operation and $\eta_{abcd}$ the
volume 4-form of the space-time.  Finally, the complex 2-forms
$\mb{U}$, $\mb{V}$, and $\mb{W}$ are given by
\begin{equation}
U_{ab}\equiv-2\ell_{[a}\bar{m}_{b]}\,, \hspace{4mm}
V_{ab}\equiv 2k_{[a}m_{b]}\,, \hspace{4mm}
W_{ab}\equiv 2m_{[a}\bar{m}_{b]}-2k_{[a}\ell_{b]} \,. \label{nu2f}
\end{equation}
Therefore, the complex scalars $\Phi_0$, $\Phi_1$, and $\Phi_2$ are
\[ \Phi_0=F_{ab}k^am^b\,,\hspace{5mm} \Phi_1=\textstyle{1\over2}
F_{ab}(k^a\ell^b+\bar{m}^am^b)\,, \hspace{5mm}
\Phi_2= F_{ab}\bar{m}^a\ell^b\,.\]
The algebraic classification of a 2-form consists of two differentiated
cases:

~

\noindent (i) The {\em regular} case ($\tilde{F}^{ab}
\tilde{F}_{ab}\neq 0$).   In this case we can pick a NP basis so that
the self-dual 2-form $\mb{\tilde{F}}$ can be written as follows
\begin{equation}
\tilde{F}_{ab} = \Phi_1 W_{ab} \,, \label{canr}
\end{equation}
where $\Phi_1$ coincide with \mbox{$-(\not\!\alpha+i\not\!\beta)$},
being \mbox{$(\not\!\alpha,-\not\!\alpha)$} and 
\mbox{$(\not\!\beta,-\not\!\beta)$} the real
eigenvalues\footnote{Here we
have changed the notation with respect to our previous paper~\cite{FASO},
where the eigenvalues where called $(\alpha,\beta)$, in order to avoid
confusion with the spin coefficients with the same name.} of $F_{ab}$
and $*F_{ab}$ respectively. Moreover, the base vectors
$\mb{\ell}$ and $\mb{k}$ are the corresponding eigenvectors.  In this adapted
basis, the 2+2 characteristic structure of regular 2-forms, sometimes called
the Maxwellian structure, appears explicitly:
$(\mb{k},\mb{\ell})$ span the 2-planes of the principal directions,
and $(\mb{m},\mb{\bar{m}})$ the orthogonal ones.  In~\cite{FASO}, a
covariant way of obtaining them for Papapetrou fields was given.

~

\noindent (ii) The {\em singular} case
($\tilde{F}^{ab}\tilde{F}_{ab}=0$).  Now we can choose the NP basis
so that $\mb{\tilde{F}}$ can be cast in the form
\begin{equation}
\tilde{F}_{ab} = \phi V_{ab}\,, \hspace{5mm}
\phi\equiv \Phi_2 \,,\label{cans}
\end{equation}
being $\mb{k}$ the only principal direction.

~
Apart from these two cases, we have the situation in which $F_{ab}=0\,.$
Obviously, in this case there are no preferred principal null
directions and the KVF is a constant vector field ($\xi_{a;b}=0$).
For vacuum spacetimes we have two different cases depending on whether
the KVF is null or non-null.  When the KVF is null the metric corresponds
to the plane-fronted gravitational waves, also called
{\em pp} waves (see, e.g.,~\cite{KSHM}).  On the other hand, when the KVF is non-null the
spacetime is Minkowski.  Therefore, since the case $F_{ab}$ is well
understood we will not consider it in what follows.

In both cases we have some freedom in the choice of the NP basis. First
we have the freedom given by the following transformations:
\begin{equation}
\mb{k} ~ \longrightarrow ~  \mb{k'} = F \mb{k}\,, \hspace{5mm}
\mb{\ell} ~ \longrightarrow ~  \mb{\ell'} = F^{-1} \mb{\ell} \,,
\label{frta}
\end{equation}
\begin{equation}
\mb{m} ~ \longrightarrow ~  \mb{m'} = \mbox{e}^{2iC}\mb{m}\,,
\label{frtb}
\end{equation}
where $F$ and $C$ are arbitrary real functions.  This exhausts the
freedom in the regular case, but in the singular case $\mb{\ell}$ is
not fixed at all, and then, we can choose its direction.  This additional
freedom is described by the following transformation:
\begin{equation}
\mb{\ell} ~ \longrightarrow ~ \mb{\ell'} = \mb{\ell} + E\mb{m}
+\bar{E}\mb{\bar{m}}+E\bar{E}\mb{k} \,, \label{frtc}
\end{equation}
where $E$ is an arbitrary complex function.  Finally, it is important
to note that whereas in the regular case the quantities
\mbox{$(\not\!\alpha,\not\!\beta)$} are invariant under the
transformations~(\ref{frta},\ref{frtb}), the quantity $\phi$ in the
singular case it is not invariant. In fact, we can use these
transformations to choose its value in an arbitrary way.

In the case of the space-time, the algebraic structure usually considered
is the algebraic structure of the Weyl tensor, which in vacuum coincides
with the Riemann tensor.   As we have said before, the algebraic
classification
of this tensor is the Petrov classification. There are different ways
of dealing with this classification (see~\cite{KSHM}), but the most
convenient one for our purposes is based on the expression for the self-dual
Weyl tensor
\begin{equation}
\tilde{C}_{abcd}=C_{abcd}+i*C_{abcd} \hspace{1cm}
(*C_{abcd}\equiv \ts{1\over2}\eta_{ab}{}^{ef}C_{efcd})
\,, \label{sdwe}
\end{equation}
in an arbitrary NP basis (see, e.g.,~\cite{KSHM})
\begin{eqnarray}
\fl \ts{1\over2}\tilde{C}_{abcd}& = \Psi_0 U_{ab}U_{cd}+
\Psi_1 (U_{ab} W_{cd} +W_{ab} U_{cd}) + \Psi_2
(V_{ab} U_{cd} +U_{ab} V_{cd} + W_{ab} W_{cd}) \nonumber \\
\fl &+\Psi_3 (V_{ab} W_{cd} +W_{ab} V_{cd}) + \Psi_4 V_{ab} V_{cd}\,,
\label{weyl}
\end{eqnarray}
Taking into account the definitions~(\ref{nu2f}) and~(\ref{weyl}), the
complex components of the Weyl tensor $\Psi_A$ ($A=0,\ldots,4$) are given by
\begin{equation}
\fl \Psi_0=\ts{1\over2}\tilde{C}_{abcd}k^am^bk^cm^d\,,\hspace{4mm}
\Psi_1=\ts{1\over2}\tilde{C}_{abcd}k^a\ell^bk^cm^d\,,\hspace{4mm}
\Psi_2=\ts{1\over2}\tilde{C}_{abcd}k^a\ell^bk^c\ell^d\,,
\label{psia}
\end{equation}
\begin{equation}
\fl \Psi_3=\ts{1\over2}\tilde{C}_{abcd}k^a\ell^b\ell^c\bar{m}^d\,,
\hspace{4mm}
\Psi_4=\ts{1\over2}\tilde{C}_{abcd}\ell^a\bar{m}^b\ell^c\bar{m}^d\,.
\label{psib}
\end{equation}
Then, we can distinguish five algebraic types (called I, II, III,
D and N) according with the number of roots, and their multiplicity,
of the polynomical equation
\[ \Psi_0+4E\Psi_1+6E^2\Psi_2+4E^3\Psi_3+E^4\Psi_4 =0 \,, \]
for the complex variable E. We will not enter here in the details (see,
e.g.,~\cite{KSHM,DINV,STEP,CHAN}), the important point is that
the complex scalars $\Psi_A$ ($A=0,\ldots,4$) contain all the information
about the algebraic structure of the space-time.  We can also include one
more case, the Petrov type O, which corresponds to conformally-flat
space-times, i.e., to a vanishing Weyl tensor ($\Psi_A=0$).

Using the algebraic structures we have just described we can introduce a
new classification of the vacuum space-times having at least one KVF, or
more precisely, of the pairs $\{(V_4,\mb{g}),\mb{\xi}\}$, which
includes the fact that there are space-times with more than one KVF.
Then, we will classify these pairs according to the following properties:
\begin{itemize}
\item The algebraic type of the Papapetrou associated with $\mb{\xi}$:
Regular (\mbox{$\not\!\alpha+i\not\!\beta$} $\neq 0$) or singular
(\mbox{$\not\!\alpha+i\not\!\beta$} $=0$).
\item The algebraic type of the space-time $(V_4,\mb{g})$, or equivalently,
of the Weyl tensor of $\mb{g}$: I, II, III, D, N, or O.
\item The degree of alignment of the principal directions of the
Papapetrou field with those of the Weyl tensor. For instance, in the
case of a singular Papapetrou field and a type N space-time there would
be only two cases, the case in which the unique principal directions of
these objects are aligned and the case in which they are not.
\end{itemize}
We can have a more refined classification by considering also some
intrinsic differential properties, specifically, the differential
properties of the principal directions both of the Papapetrou field
and the Weyl tensor: Whether they are geodesic or not, the shear,
the expansion, and the rotation.

\section{A formalism for vacuum space-times with an isometry\label{form}}
The classification presented above raises some questions, as for example
in which cases fall the known exact solutions,
or which restrictions impose each particular case of the classification,
whether or not there are empty cases, and also, whether this
classification can help in the search of new solutions.  In order to
answer these questions and to study other important related issues we are
going to introduce a new formalism.
For the sake of simplicity we will only consider here the
case of vacuum space-times $(V_4,\mb{g})$ ($R_{ab}=0$) possessing a
non-null KVF $\mb{\xi}$, i.e., its {\em norm} is different from zero
\[ N\equiv\xi^a\xi_a \neq 0 \,. \]
The case of vacuum space-times with a null Killing vector was considered in
our previous work~\cite{FASO}.  Taking into account the characteristics
of our classification, this formalism will be an extension of the
well-known NP formalism~\cite{NEPE}, which provides a clear and elegant
way of controlling the algebraic structure of the space-time. As in other
schemes where a particular basis is used, the equations and variables in
the NP formalism can be organized in the following way
(see~\cite{PAP1,PAP2} for details):
Instead of considering the
components of the metric tensor as variables we use the
components of a NP basis $(z_a{}^b)\equiv
(k^b,\ell^b,m^b,\bar{m}^b)$ with respect to a coordinate
system $\{x^a\}$. The equations for them are the expressions
defining the connection associated with such basis, which can
be found by applying the commutators of the NP basis vectors
[see equations~(\ref{comm1}-\ref{comm4}) bellow] to
the coordinate system $\{x^a\}$.  The next
set of variables are the components of this connection, which
in the NP formalism are described by the spin coefficients,
which are 12 complex scalars ($\kappa$, $\sigma$, $\rho$, $\epsilon$,
$\nu$, $\lambda$, $\mu$, $\gamma$, $\tau$, $\pi$, $\alpha$, $\beta$).
The equations for the spin coefficients (and also integrability conditions
for the previous equations) are the so-called NP equations
[see, e.g., \cite{KSHM}, equations~(7.28)-(7.45), for the sake of 
brevity we will not list these equations here], which are simply the
expressions for the Riemann tensor components in terms of the complex
connection.  In the case of vacuum space-times, the components of the
Riemann tensor are just the components of Weyl tensor, which are
considered also as new variables.  In the NP formalism they are described
by the five complex scalars $\Psi_A$ ($A=0,\ldots,4$) defined by equations
(\ref{psia},\ref{psib}).  The equations for these complex scalars come
from the second Bianchi identities [see~\cite{KSHM},
equations~(7.61)-(7.71)\footnote{Note that there is a misprint in 
equation~(7.63) of~\cite{KSHM}, instead of 
``$\ldots-\triangle\Psi_4\ldots$'' it should
read ``$\ldots-D\Psi_4\ldots$''.}, we will not list these equations here], 
which at the same time are
the integrability conditions for the NP equations.  With
these equations we get a closed system of equations for the whole
set of variables.  The literature is plenty of examples in which the
integration of these equations, in a great variety of situations, has led
to the discovery of new exact solutions of the Einstein equations
(see~\cite{KSHM} and references therein).

Now, we will extend this formalism.  The first step will be to include
new variables related to the KVF and the corresponding equations for them.
To that end, we start by writing an arbitrary KVF in a NP basis \npb 
\begin{equation}
\mb{\xi} = -\xi_l\mb{k}-\xi_k\mb{\ell}+\xi_{\bar{m}}\mb{m}+\xi_m
\mb{\bar{m}}\,, \label{kinp}
\end{equation}
where $\xi_l$, $\xi_k$, $\xi_m$, and $\xi_{\bar{m}}$ denote the components
of $\mb{\xi}$ in the NP basis, defined by
\[ \xi_k\equiv k^a\xi_a\,, \hspace{5mm}
   \xi_l\equiv \ell^a\xi_a\,, \hspace{5mm}
   \xi_m\equiv m^a\xi_a\,, \hspace{5mm}
   \xi_{\bar{m}}\equiv \bar{m}^a\xi_a\,. \]
Taking into account that $\mb{\xi}$ is a real vector field, $\xi_k$
and $\xi_l$ are real scalars, and $\xi_m$ and $\xi_{\bar{m}}$ are
complex ones and related by
\[ \xi_{\bar{m}} = \bar{\xi}_m \,. \]
Hence, we only need to consider $\xi_m$ and its complex
conjugate $\bar{\xi}_m$.  On the other hand, the norm of the KVF, $N$,
can be written in terms of $\xi_l$, $\xi_k$, and $\xi_m$ as follows
\begin{equation}
N=-2\xi_k\xi_l+2\xi_m\bar{\xi}_m\,. \label{norm}
\end{equation}

As is obvious these new variables must satisfy the Killing equations
\begin{equation}
\pounds(\mb{\xi})_{ab}=\xi_{a;b}+\xi_{b;a}=0\,, \label{kieq}
\end{equation}
where $\pounds(\mb{\xi})$ denotes Lie differentiation along $\mb{\xi}$.
Using the expression (\ref{kinp}) for the KVF, we can project the Killing
equations onto the NP basis \npb.  Then, we obtain the following
equations for the components of the Killing
\begin{equation}
D\xi_k-(\epsilon+\bar{\epsilon})\xi_k+
\bar{\kappa}\xi_m+\kappa\bar{\xi}_m=0\,, \label{kin1}
\end{equation}
\begin{equation}
\triangle\xi_l+(\gamma+\bar{\gamma})\xi_l-\nu\xi_m
-\bar{\nu}\bar{\xi}_m=0\,, \label{kin2}
\end{equation}
\begin{equation}
\triangle\xi_k+D\xi_l-(\gamma+\bar{\gamma})\xi_k
+(\epsilon+\bar{\epsilon})\xi_l+(\bar{\tau}-\pi)\xi_m+
(\tau-\bar{\pi})\bar{\xi}_m=0\,, \label{kin3}
\end{equation}
\begin{equation}
D\xi_m+\delta\xi_k-(\bar{\pi}+\bar{\alpha}+\beta)\xi_k
+\kappa\xi_l+(\bar{\rho}-\epsilon+\bar{\epsilon})\xi_m+
\sigma\bar{\xi}_m=0\,, \label{kin4}
\end{equation}
\begin{equation}
\triangle\xi_m+\delta\xi_l-\bar{\nu}\xi_k+
(\tau+\bar{\alpha}+\beta)\xi_l-(\mu+\gamma-\bar{\gamma})\xi_m-
\bar{\lambda}\bar{\xi}_m=0\,, \label{kin5}
\end{equation}
\begin{equation}
\delta\xi_m+(\bar{\alpha}-\beta)\xi_m-
\bar{\lambda}\xi_k+\sigma\xi_l=0\,, \label{kin6}
\end{equation}
\begin{equation}
\delta\bar{\xi}_m+\bar{\delta}\xi_m-(\bar{\alpha}-\beta)
\bar{\xi}_m-(\alpha-\bar{\beta})\xi_m-
(\mu+\bar{\mu})\xi_k+(\rho+\bar{\rho})\xi_l=0\,, \label{kin7}
\end{equation}
where $D$, $\triangle$, $\delta$ and $\bar{\delta}$ denote the directional 
derivatives along the NP basis vectors, defined as follows
\begin{equation} 
D\equiv k^a\partial_a \,, \hspace{5mm} 
\triangle\equiv \ell^a\partial_a \,, \hspace{5mm}
\delta\equiv m^a\partial_a \,, \hspace{5mm} 
\bar{\delta}\equiv \bar{m}^a\partial_a \,. \label{dird}
\end{equation}
Equations (\ref{kin1})-(\ref{kin7}) are completely equivalent to the Killing 
equations~(\ref{kieq}).  The usual way of dealing with the integrability
conditions for the Killing equations~(\ref{kieq}) is to
differentiate them repeatedly.  As is well-known, these integrability
conditions are equivalent to the set of equations made up of the 
equations~(\ref{rici}), which are equivalent to the equations given by
\[ \pounds(\mb{\xi}) \Gamma^a{}_{bc} = 0 \,, \]
being $\Gamma^a{}_{bc}$ the Christoffel symbols, and the following
equations
\begin{equation} 
\pounds(\mb{\xi}) R^a{}_{bcd} =
\pounds(\mb{\xi})R^a{}_{bcd; a_1;\cdots; a_N} = 0  \hspace{5mm}
(N= 1,\ldots)\,. \label{usic}
\end{equation}
As we can see, they involve derivatives of different degrees of the
Riemann tensor, the Weyl tensor in the vacuum case.

In the formalism we are developing we are going to
consider a different point of view to describe the KVF and to study
the integrability conditions.   First, we will include more equations for
the components of the KVF.  Instead of considering the Killing
equations~(\ref{kieq}) we will consider the equations that define the
Papapetrou field in terms of the KVF [equations~(\ref{papa})] which,
using the Killing equations (\ref{kieq}), can be written in the following
form
\begin{equation}
\xi_{b;a} = \textstyle{1\over2}F_{ab} \,. \label{defp}
\end{equation}
Moreover, and this is another important point, in order to write these
equations using the NP formalism we will specialize the NP basis so that
$F_{ab}$, the Papapetrou field, takes its canonical form: (\ref{canr})
in the regular case and (\ref{cans}) in the singular case.   From
equations (\ref{defp}) we will get uncoupled differential equations for
the components of the KVF.  The resulting equations can be written in
the following form, which includes both the regular and the
singular cases
\begin{equation}
 D\xi_k-(\epsilon+\bar{\epsilon})\xi_k+\bar{\kappa}\xi_m
+\kappa\bar{\xi}_m=0 \,, \label{kvf1}
\end{equation}
\begin{equation}
\triangle\xi_k-(\gamma+\bar{\gamma})\xi_k+\bar{\tau}\xi_m
+\tau\bar{\xi}_m=\textstyle{1\over2}\not\!\alpha \,, \label{kvf2}
\end{equation}
\begin{equation}
\delta\xi_k-(\bar{\alpha}+\beta)\xi_k+\bar{\rho}\xi_m
+\sigma\bar{\xi}_m=0\,, \label{kvf3}
\end{equation}
\begin{equation}
D\xi_l+(\epsilon+\bar{\epsilon})\xi_l-\pi\xi_m-\bar{\pi}\bar{\xi}_m=
-\textstyle{1\over2}\not\!\alpha \,, \label{kvf4}
\end{equation}
\begin{equation}
\triangle\xi_l+(\gamma+\bar{\gamma})\xi_l-\nu\xi_m
-\bar{\nu}\bar{\xi}_m=0\,, \label{kvf5}
\end{equation}
\begin{equation}
\delta\xi_l+(\bar{\alpha}+\beta)\xi_l-\mu\xi_m-
\bar{\lambda}\bar{\xi}_m=\textstyle{1\over2}\bar{\phi}\,, \label{kvf6}
\end{equation}
\begin{equation}
D\xi_m-(\epsilon-\bar{\epsilon})\xi_m-\bar{\pi}\xi_k
+\kappa\xi_l=0\,,  \label{kvf7}
\end{equation}
\begin{equation}
\triangle\xi_m-(\gamma-\bar{\gamma})\xi_m-\bar{\nu}\xi_k
+\tau\xi_l=-\textstyle{1\over2}\bar{\phi}\,, \label{kvf8}
\end{equation}
\begin{equation}
\delta\xi_m+(\bar{\alpha}-\beta)\xi_m-\bar{\lambda}\xi_k
+\sigma\xi_l=0 \,, \label{kvf9}
\end{equation}
\begin{equation}
\bar{\delta}\xi_m-(\alpha-\bar{\beta})\xi_m-\bar{\mu}\xi_k
+\rho\xi_l=-\textstyle{1\over2}i\not\!\beta\,. \label{kvf0}
\end{equation}
The equations for the regular case follow by putting $\phi=0$, and to
obtain the equations for the singular case we
have to put \mbox{$\not\!\alpha=\not\!\beta=0$}.
As is clear, the case \mbox{$\not\!\alpha=\not\!\beta=\phi=0$}
corresponds to the case of a constant KVF.  Furthermore, we can check 
that these equations contain the Killing equations~(\ref{kin1}-\ref{kin7}).
On the other hand, to write these equations we have used a
NP basis adapted to the Papapetrou field.  Apart from the
fact that this has allowed us to write the equations in the
simplest form, there is another important advantage related
with the classification put forward in section~\ref{alge}: It is
quite simple to implement the idea of alignment of a principal
direction of the Papapetrou field with one of the Weyl tensor.
For instance, by taking $\Psi_0=0$ we impose the principal
direction of the Papapetrou field $\mb{k}$ to be aligned with
a principal direction of the space-time.

Apart from the remarkable fact that in the set of equations
(\ref{kvf1}-\ref{kvf0}) the
directional derivatives of the components of the KVF are uncoupled,
it is important to note that we have expressions
for all the directional derivatives of all the components of the KVF.
On the other hand, these partial differential equations are linear
in the components of the KVF and inhomogeneous ($\xi_a=0$
is not a solution).  The last point is due to the appearance
of \mbox{$\not\!\alpha+i\not\!\beta$} in the regular case and $\phi$
in the singular one.  The last step in our development will be
to consider these quantities as new variables and to complete our
description by adding the corresponding equations for them.
It turns out that the equations for
\mbox{$\not\!\alpha+i\not\!\beta$} and $\phi$ are just the Maxwell
equations for the Papapetrou field, equations~(\ref{maxe}), written in
the NP basis in which it takes its canonical form [equations
(\ref{canr},\ref{cans})].   The explicit form of these equations
in the regular case is
\begin{equation}
D(\not\!\alpha+i\not\!\beta)=2\rho(\not\!\alpha+i\not\!\beta)
\,, \label{abe1}
\end{equation}
\begin{equation}
\triangle(\not\!\alpha+i\not\!\beta)=-2\mu(\not\!\alpha+i\not\!\beta)
\,, \label{abe2}
\end{equation}
\begin{equation}
\delta(\not\!\alpha+i\not\!\beta)=2\tau(\not\!\alpha+i\not\!\beta)
\,, \label{abe3}
\end{equation}
\begin{equation}
\bar{\delta}(\not\!\alpha+i\not\!\beta)=-2\pi(\not\!\alpha+i\not\!\beta)
\,, \label{abe4}
\end{equation}
and in the singular case we get 
\begin{equation}
D\phi= (\rho-2\epsilon)\phi \,, \label{phi1}
\end{equation}
\begin{equation}
\delta\phi= (\tau-2\beta)\phi \,, \label{phi2}
\end{equation}
and
\[ \kappa=\sigma=0 \,, \]
which means that the only principal direction, $\mb{k}$, is geodesic and
shear-free. 
As we can see, these equations close the system of equations for the
variables $(\xi_k,\xi_l,\xi_m)$ and \mbox{$\not\!\alpha+i\not\!\beta$}
or $\phi$.

To sum up, we have set up a formalism for vacuum space-times possessing a
KVF, which is an extension of the NP formalism, by considering the
algebraic structure associated with the KVF.  Firstly, this extension
consists in written all the equations involved in a NP basis adapted to
the algebraic structure of the Papapetrou field.  A covariant procedure
to get such a NP basis
was given in our previous paper~\cite{FASO}.  Here, we have to remember
that it is fixed up the
transformations (\ref{frta},\ref{frtb}) in the regular case and
the transformations (\ref{frta},\ref{frtb},\ref{frtc}) in the singular
case.  Secondly, we have introduced the following new variables:
(i) the components of a KVF $(\xi_k,\xi_l,\xi_m)$ in the
adapted NP basis and, (ii) a complex
variable associated with the Papapetrou field.  In the regular case it is
made up of the eigenvalues, \mbox{$\not\!\alpha+i\not\!\beta$},
whereas in the singular case it is the complex function $\phi$.
These new variables, together with the corresponding NP basis, determine
completely the Papapetrou field of $\mb{\xi}$.
The equations for the components of the KVF are (\ref{kvf1}-\ref{kvf0}),
whereas the equations for (\mbox{$\not\!\alpha$}, \mbox{$\not\!\beta$})
are (\ref{abe1}-\ref{abe4}), and the equations for $\phi$ are
(\ref{phi1},\ref{phi2}).

\section{Study of the integrability conditions\label{inte}}

In this section we will exploit the formalism we have just introduced.
The best point to start with are the equations for the
components of the KVF (\ref{kvf1}-\ref{kvf0}) because they are 
uncoupled and we know all the possible derivatives.  Then, it is
straightforward to study their integrability conditions.  To that end,  
we have to use the commutators of the directional derivatives~(\ref{dird}),
which are given by the following expressions
\begin{equation}
\triangle\,D - D\,\triangle = (\gamma+\bar{\gamma})D+
(\epsilon+\bar{\epsilon})\triangle-(\tau+\bar{\pi})\bar{\delta}-
(\bar{\tau}+\pi)\delta \,, \label{comm1}
\end{equation}
\begin{equation}
\delta\,D - D\,\delta = (\bar{\alpha}+\beta-\bar{\pi})D
+\kappa\triangle-\sigma\bar{\delta}-(\bar{\rho}+\epsilon-
\bar{\epsilon})\delta \,, \label{comm2}
\end{equation}
\begin{equation}
\delta\,\triangle - \triangle\,\delta =-\bar{\nu}D+
(\tau-\bar{\alpha}- \beta)\triangle+\bar{\lambda}\bar{\delta}
+(\mu-\gamma+ \bar{\gamma})\delta \,, \label{comm3}
\end{equation}
\begin{equation}
\bar{\delta}\,\delta - \delta\,\bar{\delta} = (\bar{\mu}-
\mu)D+(\bar{\rho}-\rho)\triangle-(\bar{\alpha}-\beta)\bar{\delta}
-(\bar{\beta}-\alpha)\delta \,. \label{comm4}
\end{equation}
Then,  the integrability conditions are obtained by applying these
commutators to the components of the KVF, $(\xi_k,\xi_l,\xi_m)$.
In this process, derivatives of the spin coefficients and derivatives
of the quantities \mbox{$\not\!\alpha+i\not\!\beta$} and $\phi$ will appear.  
We can use the NP equations [equations~(7.28-7.45) in~\cite{KSHM}] 
and the Maxwell equations [equations~(\ref{abe1}-\ref{phi2})] to eliminate 
some of them.   We have carried out this study and we have found
that the integrability conditions are local {\em algebraic} relationships
involving the complex scalars $\Psi_A$, the spin coefficients, the 
components of the KVF, and the quantities \mbox{$\not\!\alpha+i\not\!\beta$} 
and $\phi$.  This contrasts with the usual treatment where, as we have
explained before, the integrability conditions involve derivatives of
the curvature [see equations~(\ref{usic})].
In the general case, the explicit expressions for the
integrability conditions of the equations~(\ref{kvf1}-\ref{kvf0}) are
\begin{equation}
2(\Psi_0\bar{\xi}_m-\Psi_1\xi_k)= \kappa(\not\!\alpha+i\not\!\beta)
\,, \label{int1}
\end{equation}
\begin{equation}
2(\Psi_0\xi_l-\Psi_1\xi_m) = \sigma(\not\!\alpha+i\not\!\beta)
\,, \label{int2}
\end{equation}
\begin{equation}
2(\Psi_1\bar{\xi}_m-\Psi_2\xi_k) = \rho(\not\!\alpha+i\not\!\beta)-
\kappa\phi+\bar{\kappa}\bar{\phi}\,,
\label{int3}
\end{equation}
\begin{equation}
2(\Psi_1\xi_l-\Psi_2\xi_m) = \tau(\not\!\alpha+i\not\!\beta)-
\sigma\phi \,, \label{int4}
\end{equation}
\begin{equation}
2(\Psi_3\xi_k-\Psi_2\bar{\xi}_m)= \pi(\not\!\alpha+i\not\!\beta)+
\rho\phi+\bar{\sigma}\bar{\phi} \,, \label{int5}
\end{equation}
\begin{equation}
2(\Psi_3\xi_m-\Psi_2\xi_l) = \mu(\not\!\alpha+i\not\!\beta)+
\tau\phi \,, \label{int6}
\end{equation}
\begin{equation}
2(\Psi_4\xi_k-\Psi_3\bar{\xi}_m) = \lambda(\not\!\alpha+i\not\!\beta)
+\bar{\delta}\phi+2\alpha\phi \,. \label{int7}
\end{equation}
\begin{equation}
2(\Psi_4\xi_m-\Psi_3\xi_l) = \nu(\not\!\alpha+i\not\!\beta)+
\triangle\phi+2\gamma\phi\,, \label{int8}
\end{equation}
As before, the equations for the regular case are obtained by
taking $\phi=0$, and in the singular case by taking
\mbox{$\not\!\alpha+i\!\not\!\beta=0$}.  
We can see these expressions as equations for the complex scalars
of the Weyl tensor, $\Psi_A$.  Then, analyzing their structure 
we realize that they can be grouped into four pairs of equations: 
(\ref{int1},\ref{int2}) for $(\Psi_0,\Psi_1)$, 
(\ref{int3},\ref{int4}) for $(\Psi_1,\Psi_2)$, 
(\ref{int5},\ref{int6}) for $(\Psi_2,\Psi_3)$, 
and (\ref{int7},\ref{int8}) for $(\Psi_3,\Psi_4)$. 
The important point is that for each pair of equations, considered as 
equations for the two different complex scalars $\Psi_A$ that appear,
its determinant is proportional (with a non-zero proportional factor) 
to the quantity $-2\xi_k\xi_l+2\xi_m\bar{\xi}_m$, that is to say,
to the norm of the KVF~(\ref{norm}), which we have assumed to be 
non-zero.  Hence, we can solve these four pair of equations for the
corresponding Weyl complex scalars.  
Then, we get one expression for the scalars $\Psi_0$ and $\Psi_4$, 
and two expressions for the scalars $\Psi_1$, $\Psi_2$, and $\Psi_3$:
\begin{equation}
\fl \Psi_0 = \frac{\not\!\alpha+i\not\!\beta}{N}(\kappa\xi_m
-\sigma\xi_k) \,,  \label{exp0}
\end{equation}
\begin{equation}
\fl \Psi_1 = \frac{\not\!\alpha+i\not\!\beta}{N}(\kappa\xi_l
-\sigma\bar{\xi}_m) = \frac{\not\!\alpha+i\not\!\beta}{N}(
\rho\xi_m-\tau\xi_k)+\frac{\phi}{N}(\sigma\xi_k-\kappa\xi_m)
+\frac{\bar{\phi}}{N}\bar{\kappa}\xi_m \,, \label{exp1}
\end{equation}
\begin{eqnarray}
\fl \Psi_2 & = &\frac{\not\!\alpha+i\not\!\beta}{N}(\rho\xi_l
-\tau\bar{\xi}_m)+\frac{\phi}{N}(\sigma\bar{\xi}_m-\kappa
\xi_l)+\frac{\bar{\phi}}{N}\bar{\kappa}\xi_l \nonumber \\
\fl & = & \frac{\not\!\alpha+i\not\!\beta}{N}(\mu\xi_k-\pi\xi_m)+
\frac{\phi}{N}(\tau\xi_k-\rho\xi_m)-\frac{\bar{\phi}}{N}
\bar{\sigma}\xi_m \,,  \label{exp2}
\end{eqnarray}
\begin{eqnarray}
\fl \Psi_3 & = & \frac{\not\!\alpha+i\not\!\beta}{N}(\mu\bar{\xi}_m
-\pi\xi_l)+\frac{\phi}{N}(\tau\bar{\xi}_m-\rho\xi_l)
-\frac{\bar{\phi}}{N}\bar{\sigma}\xi_l \nonumber \\
\fl & = & \frac{\not\!\alpha+i\not\!\beta}{N}(\nu\xi_k-\lambda\xi_m)+
2\frac{\phi}{N}(\gamma\xi_k-\alpha\xi_m)+\frac{1}{N}
(\xi_k\triangle\phi-\xi_m\bar{\delta}\phi) \,, \label{exp3}
\end{eqnarray}
\begin{equation}
\fl \Psi_4 = \frac{\not\!\alpha+i\not\!\beta}{N}(\nu\bar{\xi}_m
-\lambda\xi_l)+2\frac{\phi}{N}(\gamma\bar{\xi}_m-\alpha\xi_l)
+\frac{1}{N}(\bar{\xi}_m\triangle\phi-\xi_l\bar{\delta}\phi) \,.
\label{exp4}
\end{equation}
These expressions constitute the first important achievement of the
proposed formalism.
They determine completely the components of Weyl tensor 
in terms of spin coefficients and quantities constructed from the KVF, 
and the dependence on these quantities is algebraic.   Therefore, they 
save us to solve the second Bianchi identities, which are the equations 
for the complex scalars $\Psi_A$.  Instead of that, we only have to 
substitute the expressions we have just obtained in the second Bianchi 
identities to obtain a set of consistency relations.  Their form will 
be explained later.  Now, it is important to remark the usefulness of 
the equations~(\ref{exp1}-\ref{exp4}), first because they can be applied 
to the question we mention before of establishing connections between
Killing symmetries and the algebraic structure of the space-time, and
second, because they can applied to any problem involving a KVF and
which requires to solve the second Bianchi identities.

The following step is to see how we can exploit this information.  
The first step is to profit the fact that we have two for expressions
for $\Psi_1$, $\Psi_2$, and $\Psi_3$.  From them we obtain three 
relationships between the components of the KVF, spin coefficients, and
the quantities \mbox{$\not\!\alpha+i\not\!\beta$} or $\phi$.  The 
explicit form of these relations is
\begin{equation}
\fl (-\tau\xi_k -\kappa\xi_l+\rho\xi_m+\sigma\bar{\xi}_m)
(\not\!\alpha+i\not\!\beta)= (\kappa\xi_m-\sigma\xi_k)\phi-
\bar{\kappa}\xi_m\bar{\phi} \,, \label{dup1}
\end{equation}
\begin{equation}
\fl (\mu\xi_k-\rho\xi_l-\pi\xi_m+\tau\bar{\xi}_m)
(\not\!\alpha+i\not\!\beta)= (-\tau\xi_k-\kappa\xi_l+\rho\xi_m+
\sigma\bar{\xi}_m)\phi + (\bar{\kappa}\xi_l+\bar{\sigma}\xi_m)
\bar{\phi} \,, \label{dup2}
\end{equation}
\begin{eqnarray}
\fl (\nu\xi_k+\pi\xi_l-\lambda\xi_m-\mu\bar{\xi}_m)
(\not\!\alpha+i\not\!\beta)= -(2\gamma\xi_k+\rho\xi_l-2\alpha\xi_m
-\tau\bar{\xi}_m)\phi \nonumber \\
\fl - \bar{\sigma}\xi_l\bar{\phi}
-\xi_k\triangle\phi + \xi_m\bar{\delta}\phi \,. \label{dup3}
\end{eqnarray}
In the regular case ($\phi=0$ and \mbox{$\not\!\alpha+i\not\!\beta\neq 0$}) 
these relations tell us that the following three complex vector fields 
must be orthogonal to the KVF $\mb{\xi}$
\[ \mb{X_1} = -\tau\mb{k}-\kappa\mb{\ell}+\rho\mb{m}+
\sigma\mb{\bar{m}} \,, \]
\[ \mb{X_2} = \mu\mb{k}-\rho\mb{\ell}-\pi\mb{m}+\tau\mb{\bar{m}}\,,\]
\[ \mb{X_3} = \nu\mb{k}+\pi\mb{\ell}-\lambda\mb{m}-\mu\mb{\bar{m}} \,.\]
It turns out that $\mb{X_1}$ and $\mb{X_3}$ are connection 1-forms in
the complex vectorial formalism of Cahen, Deveber and Defrise~\cite{CADD}
(see, e.g.,~\cite{FAFJ}).  Moreover, it is possible to see that $\mb{X_2}$
can be written as the exterior derivative of a complex function, namely
\[ \mb{X_2} = \textstyle{1\over2}\mb{d}\log(\not\!\alpha+i\not\!\beta)
\,. \]
The consequences for the singular case will be treated in the next section,
where all the possible metrics and KVFs with a singular Papapetrou
field will be determined.

Now, in the same way that we have dealt with the components of the KVF,
we will study the integrability conditions for the Maxwell equations,
that is to say, the integrability conditions for 
\mbox{$\not\!\alpha+i\not\!\beta$} in the regular case and for 
$\phi$ in the singular case.  In the regular case the situation is 
like in the case of the components of the KVF because we
know explicitly all the directional derivatives of 
\mbox{$\not\!\alpha+i\not\!\beta$}.  Then, applying
the commutators (\ref{comm1}-\ref{comm4}) to the equations
(\ref{abe1}-\ref{abe4}) and using the NP equations we can write the
integrability conditions in the regular case 
(with \mbox{$\not\!\alpha+i\not\!\beta\neq 0$}) as follows
\begin{equation} 
\delta\pi+\bar{\delta}\tau = (\bar{\alpha}-\beta)\pi+
(\alpha-\bar{\beta})\tau + \rho\bar{\mu}-\bar{\rho}\mu \,, \label{emi1}
\end{equation}
\begin{equation} 
\triangle\kappa-\bar{\delta}\sigma = \kappa(2\mu-\bar{\mu}
+3\gamma+\bar{\gamma})+\sigma(\bar{\beta}-3\alpha-2\pi-\bar{\tau})-
2\Psi_1 \,, \label{emi2}
\end{equation}
\begin{equation} 
D\pi+\bar{\delta}\rho = \rho(\alpha+\bar{\beta})-\pi(\epsilon-
\bar{\epsilon})-\bar{\kappa}\mu-\bar{\sigma}\tau \,, \label{emi3}
\end{equation}
\begin{equation} 
\triangle\tau+\delta\mu = -\mu(\bar{\alpha}+\beta)+\tau(\gamma-
\bar{\gamma})+\rho\bar{\nu}+\bar{\lambda}\pi \,, \label{emi4}
\end{equation}
\begin{equation} 
\triangle\pi-\delta\lambda = -\pi(\mu+\gamma-\bar{\gamma})-
\lambda(\bar{\alpha}-3\beta+\tau)+(\bar{\rho}-2\rho)\nu-\mu\bar{\tau}+
\Psi_3 \,. \label{emi5}
\end{equation}
As we can see, these equations only involve spin coefficients, 
in the same as the NP equations, therefore we have got more equations
for the spin coefficients.  Obviously, these integrability conditions
can have other alternative but equivalent forms by using the NP
equations.  In the singular case the situation is a little bit 
different because we only know two directional derivatives of $\phi$,
given by equations (\ref{phi1},\ref{phi2}).  We can study their
compatibility by applying the commutator (\ref{abe2}) to $\phi$.  
The result is that they are always compatible. 

We finish with some comments on the consistency of the
expressions for the Weyl complex scalars~(\ref{exp0}-\ref{exp4}) 
with respect to the second Bianchi identities in the regular case
(the singular case will be completely solved in the next section).  
As we have said before, we must insert these expressions in the second 
Bianchi identities to see what conditions they impose.  It is clear
that when we introduce the $\Psi_A$ in the second Bianchi identities 
we get expressions containing directional derivatives of the spin 
coefficients, of the components of the KVF, and of the eigenvalues 
\mbox{$\not\!\alpha$} and \mbox{$\not\!\beta$}.  Since we know all 
the directional derivatives of the last two sets of variables, they
are given in~(\ref{kvf1}-\ref{kvf0}) and~(\ref{abe1}-\ref{abe4}), they 
can be eliminated.  Therefore, the consistency conditions imposed by the 
second Bianchi identities can be seen as differential equations for the 
spin coefficients.   Due to the size of these equations we will not
write their explicit expressions here.   The conclusion of this 
discussion is that all the integrability, compatibility or consistency
conditions can be reduce to differential equations for the spin coefficients.
To sum up,  the equations for the spin coefficients that we would get
come from: (i) The NP equations (see, e.g.,~\cite{KSHM}). (ii) 
The integrability conditions for the Maxwell equations 
[equations~(\ref{emi1}-\ref{emi5})].  (iii) The equations we would 
get by introducing the expressions~(\ref{exp0}-\ref{exp4}) for
the complex scalars $\Psi_A$ in the second Bianchi identities.    
The next step in this study would be to look at the compatibility 
conditions for the whole set of differential equations for the spin 
coefficients. In the next section we study two examples in which ways of 
how to proceed further are shown.

\section{Two illustrative examples\label{exam}}

In order to show how the formalism we have just described works,
in this section we will study two particular cases of the
classification given in section~\ref{alge}: (i) Singular Papapetrou
fields in vacuum space-times.  (ii) Regular Papapetrou fields in Petrov
type III vacuum space-times.

\subsection{Singular Papapetrou fields}

In what follows we will study completely the singular case, i.e., we will
determine all the vacuum metrics admitting a KVF whose associated
Papapetrou field is singular.  We will also determine the possible KVFs 
for each case.  The singular case is characterized
by the following relations
\[ \not\!\alpha+i\not\!\beta=0 ~~ \mbox{and} ~~ \phi\neq 0 \,. \]
We start with the analysis of the integrability conditions for the 
components of the KVF, equations~(\ref{kvf1}-\ref{kvf0}), which are the 
expressions for the Weyl complex scalars $\Psi_A$, 
equations~(\ref{exp0}-\ref{exp4}).  From~(\ref{exp0}) and~(\ref{exp1}) we 
have
\begin{equation} 
\Psi_0 = \Psi_1= 0 \,, \label{ales}
\end{equation}
and therefore, the space-time must be algebraically special (this result
was already found in~\cite{FASO}, theorem 3).  Then, the Goldberg-Sachs
theorem~\cite{GOSA} tells us that $\mb{k}$ must be geodesic and shearfree
(see also~\cite{DEB2,MACI,CAMS})
\begin{equation}
\kappa = \sigma = 0 \,. \label{kasi}
\end{equation}
Another way of arriving at this result is to use the Mariot-Robinson 
theorem~\cite{MARI,ROBI}, which tells us that a vacuum space-time
containing a singular 2-form solution of Maxwell's equations (without
electromagnetic sources) must be algebraically special, and the 
principal direction of this 2-form must be geodesic and shear-free,
that is, equations (\ref{ales},\ref{kasi}) must hold in that case. 
Finally, we showed before that~(\ref{kasi}) is also a consequence of the
Maxwell equations for the Papapetrou field.  On the other hand, using 
this result we deduce, from equation~(\ref{exp2}), that
\[ \Psi_2 = 0 \,, \]
and therefore, taking into account the Petrov classification,
we have shown the following result: ``{\em The algebraic type of any 
vacuum space-time possessing a KVF such that its associated Papapetrou 
field is singular must be III, or N, or O. Moreover, the principal
direction of the Papapetrou field must be aligned with the multiple
principal direction of the space-time.}"

Following with the study, the consequences of the
equations~(\ref{dup1}-\ref{dup3}) for the singular case are
\begin{equation}
\tau\xi_k-\rho\xi_m = 0 \,,  \label{sin1}
\end{equation}
\begin{equation}
(\rho\xi_l-\tau\bar{\xi}_m)\phi+\xi_k(\triangle\phi+2\gamma\phi)
- \xi_m(\bar{\delta}\phi+2\alpha\phi) = 0 \,.\label{c2cs}
\end{equation}
At this point, we are going to use the freedom in the choice of the
NP basis in order to simplify the problem.  To that end, it is crucial
to consider two differentiated cases depending on whether $\rho$ is
zero or not.

\subsubsection{Case $\rho\neq 0$.}
In this case we can use the freedom given by the
transformation~(\ref{frtc}) to have
\[ \tau=0 \,. \]
Looking at the equations giving the change of the spin coefficients
(see, e.g., \cite{KSHM}) this can be achieved by choosing the function $E$
in~(\ref{frtc}) equals to $-\tau/\rho$.
After using this transformation, equation (\ref{sin1}) implies that
\begin{equation}
\xi_m = 0 \,. \label{cho1}
\end{equation}
Moreover, we can use the freedom (\ref{frta}) to have
\begin{equation}
\xi_l = s\xi_k\,, \hspace{5mm} s^2=1 \,. \label{cho2}
\end{equation}
Then, taking into account that now
\begin{equation}
N=-2s\xi^2_k \,, \label{nosi}
\end{equation}
the KVF will be timelike (spacelike) when $s=1$ ($s=-1$).
From the equations for the components of the KVF (\ref{kvf1}-\ref{kvf0}),
and using (\ref{cho1},\ref{cho2}), we extract the following consequences 
\begin{equation}
\epsilon+\bar{\epsilon} = \gamma+\bar{\gamma} = 0 \,, \label{afip}
\end{equation}
\[ \pi = \lambda = \rho-s\bar{\mu} = 0 \,, \]
\begin{equation}
\phi = 4s(\alpha+\bar{\beta})\xi_k = 2\nu\xi_k  \hspace{4mm}
\Longrightarrow \hspace{4mm} \nu=2s(\alpha+\bar{\beta}) \,, \label{ephi}
\end{equation}
where we have used that $\xi_k\neq 0$ [otherwise, from~(\ref{nosi}),
$N=0$]. The first part of equation
(\ref{afip}) tells us that $\mb{k}$ is affinelly parametrized. Using
the remaining freedom in the choice of the NP basis, described by
the transformation (\ref{frtb}), we can have
\[ \epsilon = \gamma = 0 \,, \]
and therefore, since $\kappa=\epsilon=\pi=0$, the NP basis is parallelly
transported along $\mb{k}$.

Introducing (\ref{ephi}) into (\ref{int5}) and taking into account
that $\xi_k$ cannot be zero, we obtain an expression
for $\Psi_3$
\begin{equation}
\Psi_3 = 2s\rho(\alpha+\bar{\beta}) \,. \label{oep3}
\end{equation}
Moreover, if we introduce (\ref{ephi}) into the Maxwell equation
(\ref{phi2}) and use the following equation
\[ \delta(\alpha+\bar{\beta}) = 0 \,, \]
which comes from the NP equation (7.41) in~\cite{KSHM} and the expression 
for $\nu$~(\ref{ephi}), we get the condition
\begin{equation}
(\alpha+\bar{\beta})(\bar{\alpha}+3\beta)\xi_k = 0 \,. \label{cocl}
\end{equation}
Therefore, there are two possibilities.  The first one is 
$\alpha+\bar{\beta}=0$,
which implies, through (\ref{oep3}) and (\ref{int7}), that the space-time 
must be Minkowski and $\phi=0$.  Moreover, we can see that these conditions 
imply $\rho=0$ against our initial assumptions. Then, we must follow the 
other possibility in (\ref{cocl}), 
i.e., we have to take
\[ \alpha = -3\bar{\beta} \,. \]
At this point, the only independent quantities
are: $\rho$, $\beta$, $\xi_k$, and $\Psi_4$.  The rest are identically
zero or can be expressed in terms of them.
From the expressions that we have obtained and from the NP equations, 
the equations for $\beta$ can be written as follows
\begin{eqnarray}
 D\beta=\bar{\rho}\beta \,, \label{dbet} \\
\triangle\beta=-s\bar{\rho}\beta \,, \nonumber \\
\delta\beta=-\textstyle{1\over4}s\bar{\Psi}_4 \,, \label{debe} \\
\bar{\delta}\beta= 0  \nonumber \,.
\end{eqnarray}
Now, let us study the compatibility of the equations~(\ref{dbet})
and~(\ref{debe}).  To that end, we have to consider the Bianchi identity
(7.63) in~\cite{KSHM}.  In our case this equation reads
\[ D\Psi_4 = 2\rho(12s\bar{\beta}^2+\Psi_4)-
4s\bar{\beta}\bar{\delta}\rho\,. \]
Then, applying the commutator (\ref{comm2}) to $\beta$, the compatibility
condition we get is
\[ \bar{\rho}\beta^2 = 0 \,. \]
Since we assumed that $\rho\neq 0$, the only possibility is $\beta=0$,
but this, as before, implies that $\alpha=\phi=\Psi_3=\Psi_4=0$, and
this implies $\rho=0$, therefore, we have reached again a contradiction.
Therefore, the conclusion is that the case $\rho\neq 0$ is empty.

\subsubsection{Case $\rho=0$.}
In this case, $\kappa=\sigma=\rho=0$ and hence, the metric belongs
to the Kundt class~\cite{KUND,KUN2,KUTR}.  As we can see in
reference~\cite{KSHM} (Chapter 27, section 27.5.1), there are only two
possible families of solutions for the Petrov types III and N. In both 
of them the line element has the form
\begin{equation}
ds^2 = 2d\zeta d\bar{\zeta} - 2du(dv+Wd\zeta+\bar{W}d\bar{\zeta}+
Hdu) \,, \label{dsku}
\end{equation}
where the metric functions are given by:

~

\noindent{\em Family 1}:
\begin{equation}
W=W^o(u,\bar{\zeta})\,,\hspace{5mm} H=\textstyle{1\over2}
(W_{,\bar{\zeta}}+\bar{W}_{,\zeta})v+H^o \,, \label{ef11}
\end{equation}
\begin{equation}
H^o=H^o(u,\zeta,\bar{\zeta}) ~~ \mbox{and} ~~
H^o{}_{,\zeta\bar{\zeta}}-\mbox{Re}(W_{,\bar{\zeta}}{}^2+
WW_{,\bar{\zeta}\bar{\zeta}}+W_{,\bar{\zeta}u})=0 \,. \label{ef12}
\end{equation}

~

\noindent{\em Family 2}:
\begin{equation}
W=-\frac{2v}{\zeta+\bar{\zeta}}+W^o(u,\zeta)\,, \hspace{5mm}
H=-\frac{v^2}{(\zeta+\bar{\zeta})^2}+\frac{W^o+\bar{W}^o}
{\zeta+\bar{\zeta}}v+H^o \,, \label{ef21}
\end{equation}
\begin{equation}
H^o=H^o(u,\zeta,\bar{\zeta}) ~~ \mbox{and} ~~
(\zeta+\bar{\zeta})\left(\frac{H^o+W^o\bar{W}^o}{\zeta+\bar{\zeta}}
\right)_{,\zeta\bar{\zeta}}=W^o{}_{,\zeta}\bar{W}^o{}_{,\bar{\zeta}}
\,. \label{ef22}
\end{equation}

~

The NP basis can be constructed as follows: $\mb{k}$ is aligned
with the principal direction of the space-time, so we can take
it to be
\begin{equation}
\mb{k} = -\mb{du}\,, \hspace{5mm} \mb{\vec{k}}=\frac{\mb{\partial}}
{\mb{\partial v}} \,. \label{npk1}
\end{equation}
Using the freedom (\ref{frtc}) we can choose $\mb{\ell}$ to be
\begin{equation}
\mb{\ell} = -(H\mb{du}+\mb{dv}+W\mb{d\zeta}+\bar{W}\mb{d\bar{\zeta}})
\,, \hspace{5mm}
\mb{\vec{\ell}} = \frac{\mb{\partial}}{\mb{\partial u}}-H
\frac{\mb{\partial}}{\mb{\partial v}} \,, \label{npk2}
\end{equation}
and finally, we can take $\mb{m}$ as follows
\begin{equation}
\mb{m} = -\mb{d\bar{\zeta}} \,, \hspace{5mm}
\mb{\vec{m}} = W\frac{\mb{\partial}}{\mb{\partial u}}-
\frac{\mb{\partial}}{\mb{\partial \zeta}}\,. \label{npk3}
\end{equation}
Then, using the explicit expressions given in (\ref{ef11}-\ref{ef22}) for
each family we can study whether or not there are solutions and to
determine them.  To that end we are going to consider the Petrov types
III and N (the only possible ones) separately.

In the case of Petrov type III solutions ($\Psi_3\neq 0$), from
equation (\ref{int5}) we have
\[ \xi_k = 0 \,.\]
Taking into account this fact, we have studied the consequences of 
the equations (\ref{kvf1}-\ref{kvf0}) using the expressions given in 
the~\ref{appa} for the spin coefficients and the scalars $\Psi_3$ and 
$\Psi_4$ in each family.  In this study we have assumed, according 
to the hypothesis that define this subcase, that $\phi\neq 0$ and 
$\Psi_3 \neq 0$.  After some calculations we have found that these 
equations are incompatible.  Therefore, the conclusion is that 
{\em there is not any Petrov type III vacuum solution with a KVF 
having a singular Papapetrou field.}  Hence, only Petrov N and O 
vacuum space-times can have a KVF with a singular Papapetrou field.

Now, let us study the case of Petrov type N solutions, characterized
by $\Psi_3=0$.  In this situation, equation~(\ref{int4}) implies that
\[ \tau = 0 \,. \]
This condition is very important because it is invariant under any
of the freedoms we have in the choice of the NP basis
[transformations~(\ref{frta},\ref{frtb},\ref{frtc})].  If we look 
now at the expressions given in~\ref{appa} for the spin coefficients,
we can see that $\tau\neq 0$ for the second family of solutions.
Therefore, we can only find solutions in the first family.
In this family the condition $\Psi_3=0$ implies
\[ W^o{}_{\bar{\zeta}\bar{\zeta}}=0 \,,\]
and using the remaining freedom in the choice of the coordinates
(see~\cite{KSHM}) we can have $W^o=0$.  The resulting line-element
corresponds to a {\em pp} wave.  All the particular classes
of these space-times admitting KVFs, apart from the null KVF
$\mb{\partial}/\mb{\partial v}$, were studied and classified by
Ehlers and Kundt~\cite{EHKU} (see also~\cite{KSHM}).   We have studied
the consequences of the equations~(\ref{kvf1}-\ref{kvf0}) and we have
found that the solutions allowed belong to the following two classes:

~

\noindent{\em Class 1}:  The only metric function, $H$, can be written 
as follows
\begin{equation}
H = f(\zeta)+\bar{f}(\bar{\zeta})\,, \label{ppw1}
\end{equation}
where $f$ is an arbitrary complex function of $\zeta$.  This case was 
already studied in~\cite{FASO}. The KVF is given by
\begin{equation}
\fl \mb{\vec{\xi}} = \frac{\mb{\partial}}{\mb{\partial u}}\,,
\hspace{5mm} \mb{\xi} = -\mb{dv}-2H\mb{du} ~~ \Longleftrightarrow ~~
\xi_k=-1\,, \hspace{5mm} \xi_l=-H\,,\hspace{5mm} \xi_m=0\,, \label{kpp1}
\end{equation}
and hence, $N=-2H$. Therefore, the KVF can be either timelike or
spacelike.  Its associated Papapetrou field is given by
\[ \mb{F} = 2\mb{du}\wedge(H_{,\zeta}\mb{d\zeta}+
H_{,\bar{\zeta}}\mb{d\bar{\zeta}}) ~~ \Longrightarrow ~~
\phi= 2H_{,\bar{\zeta}} \,. \]

~

\noindent{\em Class 2}: Now, the metric function $H$ has the
following form
\begin{equation}
H = f(u,\zeta)+\bar{f}(u,\bar{\zeta})\,, \hspace{5mm}
f(u,\zeta)= A(u)\zeta^2 \,, \label{ppw2}
\end{equation}
where $A$ is an arbitrary complex function of $u$. In this case we
have
\[ \xi_k=0\,,\hspace{5mm} \xi_m=\xi_m(u)\,, \hspace{5mm}
\xi_l=\xi_m\zeta+\bar{\xi}_m \bar{\zeta} \,, \]
therefore, the KVF is spacelike and is given by
\begin{equation}
\fl \mb{\vec{\xi}}=-(\xi'_m\zeta+\bar{\xi}'_m\bar{\zeta})
\frac{\mb{\partial}}{\mb{\partial v}}-\bar{\xi}_m
\frac{\mb{\partial}}{\mb{\partial\zeta}}
-\xi_m\frac{\mb{\partial}}{\mb{\partial\bar{\zeta}}}\,,\hspace{5mm}
\mb{\xi}=(\xi'_m\zeta+\bar{\xi}'_m\bar{\zeta})\mb{du}-
\bar{\xi}_m\mb{d\bar{\zeta}}-\xi_m\mb{d\zeta}\,, \label{kpp2}
\end{equation}
where $'\equiv d/du$.  From this expressions, the Papapetrou field
is 
\[ \mb{F}= -2\mb{du}\wedge(\xi'_m\mb{d\zeta}+\bar{\xi}'_m
\mb{d\bar{\zeta}}) ~~ \Longrightarrow ~~ \phi=-2\bar{\xi}'_m \,.\]

~

In both classes of solutions the principal direction is
$\mb{\partial}/\mb{\partial v}$ [see equation~(\ref{npk1})], that is,
it is parallel to the null KVF (the normal to the wave fronts) and
also to the principal direction of the space-time.

To complete the study of the singular case we have to consider the
Petrov type O.  Since we are considering vacuum space-times the only
possibility is the Minkowski space-time
\begin{equation}
ds^2= -2 du dv+2d\zeta d\bar{\zeta} \,. \label{mink}
\end{equation}
A very convenient NP basis is the following
\[ \fl \mb{\vec{k}}=\frac{\mb{\partial}}{\mb{\partial v}}\,,\hspace{4mm}
\mb{\vec{\ell}}=\frac{\mb{\partial}}{\mb{\partial u}}\,,\hspace{4mm}
\mb{\vec{m}}=\frac{\mb{\partial}}{\mb{\partial \zeta}}\,,\hspace{10mm}
\mb{k}=-\mb{du}\,,\hspace{4mm} \mb{\ell}=-\mb{dv}\,,\hspace{4mm}
\mb{m}=\mb{d\bar{\zeta}} \,. \]
All the spin coefficients associated with this basis vanish
and then it is simple to find all the possible KVFs.  The most general
KVF having a singular Papapetrou field is given by
\[ \xi_k = C_1\,,\hspace{5mm} \xi_l=\textstyle{1\over2}(\bar{\phi}
\zeta+\phi\bar{\zeta}+C_2)\,,\hspace{5mm}
\xi_m=-\textstyle{1\over2}(\bar{\phi}u+C_3) \,,\]
\begin{equation}
\fl \mb{\vec{\xi}}=-\textstyle{1\over2}(\bar{\phi}\zeta+\phi\bar{\zeta}+C_2)
\frac{\mb{\partial}}{\mb{\partial v}}
-C_1\frac{\mb{\partial}}{\mb{\partial u}}-\textstyle{1\over2}
(\phi u+\bar{C}_3)\frac{\mb{\partial}}{\mb{\partial \zeta}}-
\textstyle{1\over2}(\bar{\phi}u+C_3)
\frac{\mb{\partial}}{\mb{\partial\bar{\zeta}}}\,, \label{kvfm}
\end{equation}
where $C_1$, $C_2$ are arbitrary real constants and $C_3$ and $\phi$
are arbitrary complex constants.  Then, the Papapetrou field associated
with this KVF has the following form
\[ \mb{F} = -\mb{du}\wedge(\bar{\phi}\mb{d\zeta}+\phi
\mb{d\bar{\zeta}}) \,. \]

With this we have finished the study of the singular case, that is to say,
we have found all the possible vacuum metrics possessing a KVF with 
a singular Papapetrou field.  To sum up, we have showed the following 
result: ``{\em The only vacuum space-times admitting a KVF whose associated 
Papapetrou field is singular are the classes of pp waves given 
by~(\ref{ppw1}) and~(\ref{ppw2}) and the Minkowski space-time~(\ref{mink})}".
The corresponding KVFs are given in~(\ref{kpp1}),~(\ref{kpp2})
and~(\ref{kvfm}) respectively.

\subsection{Regular Papapetrou fields in Petrov type
III vacuum space-times}

The previous study covers a branch of the classification put forward
in section~\ref{alge}.  In what follows, we will study another particular
case of this classification in which the space-time is Petrov type III
and the Papapetrou field is regular.  Specifically, we are going to show
the following statement:  ``{\em In any Petrov type III vacuum space-time
with a Killing vector field whose Papapetrou field is regular, the multiple 
principal direction of the space-time cannot be aligned with either of the 
two principal directions of the Papapetrou field.}"

This means that the algebraic structure of the Papapetrou
field cannot be completely adapted to that of the Weyl tensor
contrary to what happens, for instance, in the case of the Kerr 
metric~\cite{KERR}, where
the two principal directions of the Papapetrou field associated
with the timelike KVF coincide with the two principal multiple
directions of the space-time (see, e.g.,~\cite{FASO}).  More precisely, 
the statement above implies that the only possibility of alignment is
between the single direction of the space-time and one of the
principal directions of the Papapetrou field. In other words, the cases
of the classification corresponding to single and double alignment
with the principal multiple direction of the space-time are empty.

The idea to prove the statement is to assume
that one of the principal directions of the Papapetrou field,
say $\mb{k}$, is aligned with the multiple principal direction of
the space-time, and then, to reach a contradiction.
As is clear, the consequences of the alignment are
\[ \Psi_0=\Psi_1=\Psi_2=0 \,.\]
In this situation, the consequences of the integrability conditions
(\ref{int1}-\ref{int8}) [or equivalenty, expressions
(\ref{exp0}-\ref{exp4})] are: First, $\mb{k}$ is geodesic and
shear-free (this is also a consequence of the Goldberg-Sachs
theorem~\cite{GOSA})
\begin{equation}
\kappa = \sigma = 0\,. \label{gesh}
\end{equation}
$\mb{k}$ is expansion- and rotation-free
\begin{equation}
\rho = 0\,, \label{rofr}
\end{equation}
the spin coefficient $\tau$ also vanishes
\begin{equation}
\tau = 0 \,, \label{tauv}
\end{equation}
and the following remaining conditions
\begin{equation}
2\Psi_3 \xi_k = \pi(\not\!\alpha+i\not\!\beta) \,, \label{t3c1}
\end{equation}
\begin{equation}
2(\Psi_3\xi_l-\Psi_4\xi_m) = -\nu(\not\!\alpha+i\not\!\beta)\,,
\label{t3c2}
\end{equation}
\begin{equation}
2\Psi_3\xi_m = \mu(\not\!\alpha+i\not\!\beta) \,, \label{t3c3}
\end{equation}
\begin{equation}
2(\Psi_3\bar{\xi}_m-\Psi_4\xi_k) = -\lambda(\not\!\alpha+i\not\!\beta)
\,. \label{t3c4}
\end{equation}
From (\ref{gesh},\ref{rofr}) we deduce that the metric must belong
to the Kundt class of solutions~\cite{KUND}.  And in particular,
since we are considering Petrov type III vacuum space-times, it must
belong to the families given in the previous subsection
[Equations~(\ref{dsku}-\ref{ef22}) and~\ref{appa}].  Moreover, we can
use the remaining freedom in the
choice of the NP basis, given by transformations (\ref{frta},\ref{frtb}),
to have
\[ \epsilon = \alpha = \beta = 0 \,. \]

The next step will be to study the stability of the relationships
(\ref{t3c1}-\ref{t3c4}) under
the action of the directional derivatives $D$, $\triangle$, $\delta$,
and $\bar{\delta}$. That is to say, we will apply these operators to
them and we will use the second Bianchi identities and the NP equations to
substitute the directional derivatives of $(\Psi_3,\Psi_4)$ and
of the spin coefficients respectively.  Finally, we will check
whether or not we get new conditions.  To that end,
we will assume that $\Psi_3\neq 0$, otherwise the space-time would
be of Petrov type N.

Applying the four directional derivatives to (\ref{t3c1}) we get
expressions for the four directional derivatives of $\pi$.  Then,
applying $D$ to (\ref{t3c2}), using the expression we have got for
$\triangle\pi$, and the equations (\ref{t3c1}-\ref{t3c4}), we arrive
at the following interesting equation (always assuming that
\mbox{$\not\!\alpha+i\not\!\beta\neq 0$})
\[ \pi\delta\Psi_4-\mu D\Psi_4+\Psi^2_3 = 0 \,. \]
This relation shows that ``{\em in Petrov III vacuum space-times 
with a KVF we cannot have a double alignment of the principal
directions of the Weyl tensor with those of the Papapetrou field.}"
Otherwise, it would imply $\Psi_4=0$, and this 
would imply that $\Psi_3=0$, which would lead to the Minkowski 
space-time, in contradiction with our assumptions.

We could follow further the study of the integrability conditions
but since we know that the metric must belong to the families
given in~(\ref{dsku}-\ref{ef22}), we can use these expressions,
like in the previous subsection, to determine whether or not
there are solutions and in the case that there were, to find their
exact form.   However, there is an important difference with
respect to the singular case: Now we do not have the freedom
in the choice of the NP basis given by~(\ref{frtc}) (the direction
determined by $\mb{\ell}$ is fixed because it corresponds now to
one of the principal directions of the Papapetrou field), and 
therefore we cannot fix the NP basis like 
in~(\ref{dsku}-\ref{ef22}).  The only exception is $\mb{k}$ because 
it must be aligned with the principal direction of the space-time 
and hence, it must coincide with~(\ref{npk1}).

On the other hand, since condition~(\ref{tauv}) holds we can use the
same argument as in the singular case, namely, independently of the
directions of $\mb{\ell}$ and $\mb{m}$, the condition~(\ref{tauv}) will
still hold (see, e.g.,~\cite{KSHM} for the formulae for the
transformation of the spin coefficients).  Therefore, taking
into account that in the second family $\tau$ cannot vanish
(see \ref{appa}) we deduce that we can only find solutions in
the first family.

For this study we can take a NP basis consisting of~(\ref{npk1}) and
\[ \mb{\ell'}=\mb{\ell}+E\mb{m}+\bar{E}\mb{\bar{m}}+E\bar{E}\mb{k}
\,, \hspace{5mm}  \mb{m'}=\mbox{e}^{2iC}\mb{m} \,, \]
where $\mb{\ell}$ and $\mb{m}$ are given in (\ref{npk2},\ref{npk3}),
and $E$ and $C$ are complex and real functions respectively that
must be determined.  In this situation, we have studied the equations
for the components of the KVF~(\ref{kvf1}-\ref{kvf0}) and we have found 
that they have not any solution for the Petrov type III.
Therefore,  this completes the proof of the statement given at the
beginning of this subsection.

\section{Remarks and conclusions\label{core}}

In this paper we have set up a formalism to study vacuum space-times 
with a KVF.  This formalism exploits the fact that we can associate
an algebraic structure with the KVF through its exterior derivative,
the Papapetrou field.  Introducing new variables related with the
Papapetrou field, and writing all the equations with respect to a NP
basis adapted to its algebraic structure, we have obtained a new
framework in which we can study the connections between the existence
of Killing symmetries and the algebraic structure of the space-time,
a subject scarcely studied in the literature.  Moreover, this
formalism provides, in a natural way, a classification of the
space-times with a KVF.  The cases in which there are alignments
of the principal direction(s) of the Papapetrou field with those
of the space-time are, {\em a priori}, the most simple to be
dealt with this formalism.  In this sense, in this paper we have
seen that in the case of singular Papapetrou fields there is
always alignment and that the class of space-times is very limited.
In the case of regular Papapetrou fields we have studied
Petrov type III vacuum space-times arriving at the conclusion that
there is no possible alignment of the multiple principal direction
of the space-time with some of the two principal directions of the
Papapetrou field.   In contrast with this situation, there are other
vacuum space-times in which we can find alignments.  An interesting
example is the case of the Kerr metric in which the two multiple
principal directions of the space-time are aligned with those of the
Papapetrou field (see, e.g.,~\cite{FASO}).  Other cases with alignment
have been studied in~\cite{FSTS}.

On the other hand, it is remarkable the fact that the study of the
integrability conditions
for the components of the KVF leads directly to explicit expressions
for the components of the Weyl tensor in terms of the connection
(spin coefficients) and components of the Papapetrou field
[(\mbox{$\not\!\alpha$},\mbox{$\not\!\beta$}) or $\phi$].  Then,
we do not need to solve the second Bianchi identities, but
to study the conditions that they impose on the spin coefficients.
In conclusion, this formalism is suitable for the study of any problem
or situation in which the knowledge of the Weyl complex scalars is
required.   Some topics in which this formalism may be helpful are:
Search and study of exact solutions,
perturbations of black holes preserving a symmetry (e.g., axisymmetric
or stationary perturbations), the question of the equivalence of metrics,
the construction of numerical algorithms, etc.

Finally, it is worthwhile to discuss the possible extensions of this
formalism.   In this sense it is important to note that the assumption
that the space-time would have a vanishing energy-momentum tensor
is not fundamental for the development of the formalism.  Therefore, a
possible way of extending it is to consider other types of energy-momentum
content.  Other kind of extension would be to consider the present
scheme applied to other symmetries, as for instance those generated
by conformal KVFs.

\ack

We wish to thank Malcolm MacCallum for pointing out to us some
references.
Some of the calculations in this paper were done using the
computer algebraic system REDUCE.
F.F. acknowledges financial support from the D.G.R. of the Generalitat de
Catalunya (grant 1998GSR00015), and the Spanish Ministry of Education
(contract PB96-0384).
C.F.S. wishes to thank the Alexander von Humboldt Foundation for financial
support and the Institute for Theoretical Physics of the Jena University
for hospitality during the first stages of this work.  Currently, C.F.S.
is supported by the European Commission (contract HPMF-CT-1999-00149).

\appendix

\section{Explicit expressions for the spin coefficients\label{appa}}
In this appendix we consider the NP basis \npb given in equations
(\ref{npk1}-\ref{npk3}), associated with the subclass of Kundt metrics
for vacuum and Petrov types III and N, which are determined by
expressions (\ref{dsku}-\ref{ef22}).  Using these expressions we can
compute the spin coefficients and the non-zero Weyl complex scalars
$\Psi_3$ and $\Psi_4$.  After some calculations, the result is

~

\noindent{\em Family 1}:
\[\fl \kappa=\epsilon=\sigma=\rho=\tau=\pi=\alpha=\beta=\lambda=0 \,,
\hspace{5mm} \gamma=\textstyle{1\over2}W_{,\bar{\zeta}}\,,\hspace{5mm}
\mu=\textstyle{1\over2}(W_{,\bar{\zeta}}-\bar{W}_{,\zeta}) \,, \]
\[\fl \nu = -H^o{}_{,\bar{\zeta}}+\bar{W}_{,u}+
\textstyle{1\over2}(\bar{W}_{,\zeta}+W_{,\bar{\zeta}})\bar{W}-
\textstyle{1\over2}vW_{,\bar{\zeta}\bar{\zeta}}\,, \]
\[\fl \Psi_3 = -\textstyle{1\over2}W_{,\bar{\zeta}\bar{\zeta}}\,, \hspace{5mm}
\Psi_4=H^o{}_{,\bar{\zeta}\bar{\zeta}}-\bar{W}
W_{,\bar{\zeta}\bar{\zeta}}+\textstyle{1\over2}v
W_{,\bar{\zeta}\bar{\zeta}\bar{\zeta}} \,. \]

\noindent{\em Family 2}:
\[\fl \kappa=\sigma=\rho=\epsilon=\lambda=0\,, \hspace{5mm}
\tau=-\pi=2\alpha=2\beta=\frac{1}{\zeta+\bar{\zeta}}\,, \hspace{5mm}
\gamma=-\frac{v}{(\zeta+\bar{\zeta})^2}+\frac{\bar{W}^o}{\zeta+\bar{\zeta}}
\,, \]
\[\fl \mu=-\frac{W^o-\bar{W}^o}{\zeta+\bar{\zeta}}\,, \hspace{5mm}
\nu = -H^o{}_{,\bar{\zeta}}+\bar{W}^o_{,u}+
\frac{2H^o-v\bar{W}^o{}_{,\bar{\zeta}}+\bar{W}^o{}^2+W^o\bar{W}^o}
{\zeta+\bar{\zeta}}+v\frac{W^o-\bar{W}^o}{(\zeta+\bar{\zeta})^2} \,,\]
\[\fl \Psi_3= -\frac{\bar{W}^o{}_{,\bar{\zeta}}}{\zeta+\bar{\zeta}}\,,\]
\[\fl \Psi_4=H^o{}_{,\bar{\zeta}\bar{\zeta}}-
\bar{W}^o{}_{,u\bar{\zeta}}+\frac{v\bar{W}^o{}_{,\bar{\zeta}\bar{\zeta}}
-2H^o{}_{,\bar{\zeta}}-(W^o+3\bar{W}^o)\bar{W}^o{}_{,\bar{\zeta}}}
{\zeta+\bar{\zeta}}+2\frac{H^o+v\bar{W}^o{}_{,\bar{\zeta}}+
W^o\bar{W}^o}{(\zeta+\bar{\zeta})^2}\,. \]



\section*{References}
\begin{thebibliography}{99}
\bibitem{KSHM} Kramer D, Stephani H,  MacCallum  M and Herlt E 1980
{\it Exact solutions of Einstein's field equations} (Berlin: VEB Deutscher
Verlag der Wissenschaften)
\bibitem{PAPA} Papapetrou A 1966 {\it Ann. Inst. H. Poincar\'e} {\bf A4}
83
\bibitem{DEB1} Debney G C 1971 \JMP {\bf 12} 1088
\bibitem{DEB2} Debney G C 1971 \JMP {\bf 12} 2372
\bibitem{HOMI} Horsk\'y J and Mitskievitch N V 1989 {\it Czech. J. Phys.}
{\bf B39} 957
\bibitem{CAMI} Cataldo M and Mitskievitch N V 1990 \JMP {\bf 31} 2425
\bibitem{WALD} Wald R M 1974 \PR {\bf D10} 1680
\bibitem{FASO} Fayos F and Sopuerta C F 1999 \CQG {\bf 16} 2965
\bibitem{NEPE} Newman E T and Penrose R 1962 \JMP {\bf 3} 566
\bibitem{PETR} Petrov A Z 1954 {\it Uch. zap. Kazan Gos. Univ.}
{\bf 114} (book 8) 55
\bibitem{BEL1} Bel L 1958 {\it C.R. Acad. Sci. Paris.} {\bf 247} 2096
\bibitem{BEL2} Bel L 1959 {\it C.R. Acad. Sci. Paris.} {\bf 248} 2561
\bibitem{DINV} d'Inverno R 1992 {\it Introducing Einstein's Relativity}
(Oxford: Oxford University Press).
\bibitem{STEP} Stephani H 1982 {\it General relativity: an introduction
to the gravitational field} (Cambridge: Cambridge University Press).
\bibitem{CHAN} Chandrasekhar S 1983 {\it The mathematical theory of
black-holes} (New York: Oxford University Press)
\bibitem{PAP1} Papapetrou A 1971 {\it C.R. Acad. Sc. Paris.} {\bf 272} 1537
\bibitem{PAP2} Papapetrou A 1971 {\it C.R. Acad. Sc. Paris.} {\bf 272} 1613
\bibitem{CADD} Cahen M, Debever R and Defrise L 1967 {\it J. Math. Mech.}
{\bf 16} 761
\bibitem{FAFJ} Fayos F, Ferrando J J and Ja\'en X 1990 \JMP {\bf 31}
410
\bibitem{GOSA} Goldberg J N and Sachs R K 1962 {\it Acta Phys. Polon.,
Suppl.} {\bf 22} 13
\bibitem{MACI} McIntosh C B G 1976 {\it Gen. Rel. Grav.} {\bf 7} 215
\bibitem{CAMS} Catenacci R, Marzuoli A and Salmistraro F 1980 {\it Gen.
Rel. Grav.} {\bf 12} 575
\bibitem{MARI} Mariot L 1954 {\it C.R. Acad. Sci. Paris.} {\bf 238} 2055
\bibitem{ROBI} Robinson I 1961 \JMP {\bf 2} 290
\bibitem{KUND} Kundt W 1961 {\it Z. Physik} {\bf 163} 77
\bibitem{KUN2} Kundt W 1962 {\it Proc. Roy. Soc. Lond.} {\bf A 270}
328
\bibitem{KUTR} Kundt W and Tr\"umper M 1962 {\it Akad. Wiss. Lit.
Mainz, Abhandl. Math.-Nat. Kl.} {\bf 12}
\bibitem{EHKU} Ehlers J and Kundt W 1962 in {\em Gravitation: an
introduction to current research} (edited by L.Witten, Wiley).
\bibitem{KERR} Kerr R P 1963 \PRL {\bf 11} 237
\bibitem{FSTS} Fayos F and Sopuerta C F 2000 {\it in preparation.}
\end{thebibliography}
\end{document}